\documentclass[runningheads]{llncs}
\usepackage{graphicx}
\begin{document}
\title{A Fast Counting Method for 6-motifs \\
with Low Connectivity
}
%
%

\author{Taha Sevim \and
Muhammet Sel\c{c}uk G\"uvel \and
Lale \"Ozkahya}
\authorrunning{T. Sevim et al.}
%
\institute{Department of Computer Engineering, Hacettepe University, Ankara, Turkey
\email{\{tahasevim231,selcukguvel,laleozkahya\}@gmail.com}}
%
\maketitle              
\begin{abstract}
A $k$-motif (or graphlet) is a subgraph on $k$ nodes in a graph or network. 
Counting of motifs in complex networks has been a well-studied problem in network analysis 
of various real-word graphs arising from the study of social networks and bioinformatics. 
In particular, the triangle counting problem has received much attention due to its 
significance in understanding the behavior of social networks. Similarly, subgraphs with more 
than 3 nodes have received much attention recently. 
While there have been successful methods developed on this problem, most of the existing algorithms 
are not scalable to large networks with millions of nodes and edges. 

The main contribution of this paper is a preliminary study 
that genaralizes the exact counting algorithm provided by Pinar, Seshadhri and Vishal to a collection of 6-motifs. 
This method uses the counts of motifs with smaller size to obtain the counts of 6-motifs 
with low connecivity, that is, containing a cut-vertex or a cut-edge. 
Therefore, it circumvents the combinatorial explosion that 
naturally arises when counting subgraphs in large networks. 

\keywords{social networks \and motif analysis \and subgraph counting.}
\end{abstract}

\section{Introduction}
In social network analysis, any fixed subgraph with $k$ nodes  
is called a {\it $k$-motif (or graphlet)} and their analysis has 
been a useful method to characterize the structure of real-world graphs. 
It has observed particularly in social networks, that some motifs are 
more common than others, and the structure of the network is different than 
the structure of the random graphs~\cite{holland1976local,milo2002,watts1998collective}.
While knowing that only the analysis of these subgraphs is not sufficient 
to understand the structure of the networks, it has been validated that the 
motif frequencies provide substantial information about the local network 
structure in various domains~\cite{holland1976local,faust2010puzzle,frank1988}. 
By counting the number of embeddings of each motif in a network, 
it is possible to create a profile of sufficient statistics 
that characterizes the network structure~\cite{shervashidze2009}.

Although there has been significant amount of success and impact on areas varying from social science to biology, the search for faster and more efficient algorithms to compute the frequencies of graph patterns continues. 
The main reason to study algorithms to count motifs faster is combinatorial explosion. The 
running time of algorithms to exactly count $k$-motifs on the vertex set $V$ 
is of the order $O(|V|^k)$. The counts of 6-motifs are in the orders 
of billions to trillions for graphs with more than a few million edges. 
Thus, an enumeration algorithm cannot terminate in a reasonable time. 
The idea presented in~\cite{escape} and extended to 6-motifs here uses a 
framework of counting with minimal enumeration. 
The main contribution of this paper is a preliminary study 
that genaralizes the exact counting algorithm provided in~\cite{escape} to a 
collection of 6-motifs.  
To the best of our knowledge, 
this is the only study that counts 6-motifs using exact computation and 
performs all counts in graphs with millions of edges in minutes.
As a preliminary work, we are able to exactly count the motifs shown in Figure~\ref{fig:6-motif}. 
The particular reason that this subset of motifs are chosen is that each of them contains 
a cut-vertex or a cut-edge, that is, removing that vertex or edge makes the motif disconnected. 
The main idea is to build a framework to cut each pattern of 6 nodes 
into smaller patterns, where each of the patterns contain that particular cutting subset, also 
called {\it cut-set.} Then, the enumeration is only needed for these smaller patterns 
rather than the big pattern. For our purposes, we do not carry out the enumeration 
and use the counts for these smaller patterns obtained in~\cite{escape}. 

There are various approximation algorithms~\cite{guise,jha2015path,graft,seshadhri2013,fanmod}, 
however the results they provide are not exact and scalable for counting larger motifs with more 
than 4 nodes, whereas the method presented here 
is also scalable to very large networks. 
As presented in Section~\ref{results}, 
our method is able to count 6-motifs in 
Figure~\ref{fig:6-motif} for a network with 3 millions of edges under 5 minutes. 
Most of the studies on counting motifs have been 
focusing on smaller motifs with size at most 4. 
In particular, the count of triangles has been widely 
studied due to its importance in the analysis of social networks~\cite{ugander}. 
These results have been helpful for graph classification and often used as graph attributes. 
Another group of recent studies on subgraph 
counts are used for detecting communities
and dense subgraphs, such as \cite{benson2016higher,sariyuce2015finding,tsourakakis2015k,tsourakakis2017scalable}. 
More recent algorithmic improvements on counting triangles can be 
found in~\cite{seshadhri2013,tsourakakis2011}. Exact and 
approximate algorithms for computing the number of non-induced 
4-motifs are proposed in~\cite{gonen2009}. 

It has been observed that sampling algorithms~\cite{guise,graft,wernicke2006,fanmod} and randomized algorithms, such as the 
color coding method~\cite{betzler2011,hormozdiari2007,sahad}, are not feasible for counting motifs 
of size larger than 4. One of the most recently developed algorithms in~\cite{motivo} 
estimates the number of 7-motifs on a graph with 65M nodes and 1.8B edges in around 40 minutes. 
Exact counting algorithms as in~\cite{kashtan,kavosh,fanmod} exist, but they are very slow and 
not scalable to large graphs. 
The recent study in~\cite{escape} showed an exact counting 
technique that counts all patterns with at most 5 vertices on graphs with tens of 
millions of edges in several minutes. 

The main contribution in this paper is to cut each pattern in the chosen collection 
into smaller patterns 
and use the enumeration on these smaller patterns to count the big pattern by using 
the framework in~\cite{escape}. Some other 
algorithms that used ideas to avoid enumeration can be seen in ~\cite{ahmed2015,elenberg2015,elenberg2016,hovcevar2014}.
\section{Methodology}
\begin{figure}
\vspace{-.5cm}
    \centering
    \includegraphics [scale=0.2]{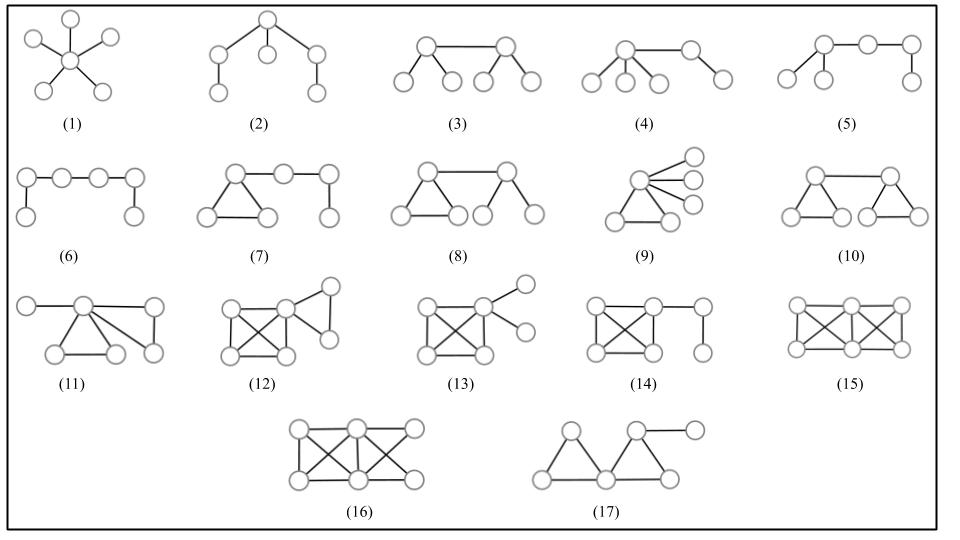}
    \caption{The 6-motifs with low connectivity.}
    \label{fig:6-motif}
    \vspace{-.5cm}
\end{figure}

The input graph $G=G(V,E)$ is undirected, where $V$ and $E$ denote the vertex set 
and the edge set of $G$, respectively. 
A subgraph of $G$ is called {\it induced} if all edges present in the host graph exist as edges in that 
subgraph. 
Otherwise, it is called {\it non-induced.} 
In our counting method, a subgraph means a non-induced subgraph. 
We call a triangle with a missing edge a {\it wedge},  
a $K_4$ with a missing edge a {\it diamond} and 
a triangle with an edge attached to one of its vertices, 
a {\it tailed triangle.} 
\begin{figure}
     \centering
     \vspace{-.5cm}
     \includegraphics[scale=0.2]{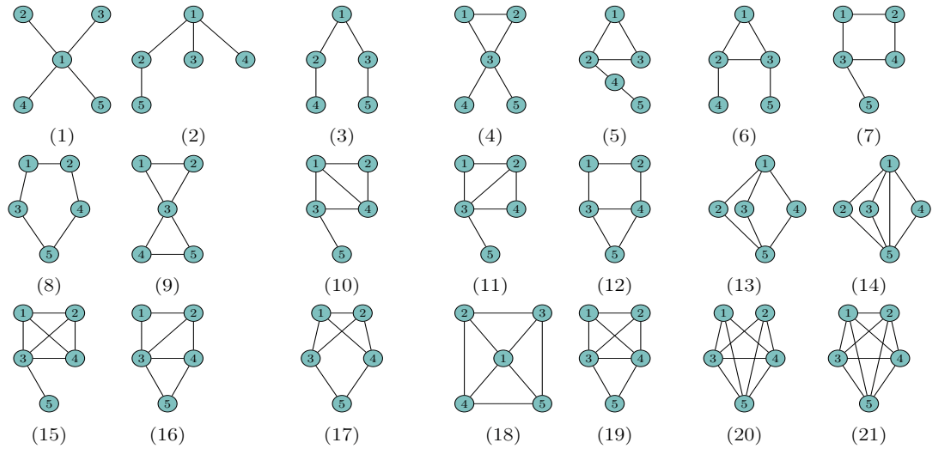}
     \caption{The collection of connected 5-motifs~\cite{escape}.}
     \label{fig:escape5}
     \vspace{-.5cm}
 \end{figure}
In our notation, for each vertex $i,$ we use $d(i)$ and $T(i),$ (resp. $T(e)$)  
to denote the degree of $i$ and the number of 
triangles that contain the vertex $i$ (resp. edge $e$), respectively. 
Similarly, $C_4(i)$ (resp. $C_4(e)$) and $K_4(i)$ (resp. $K_4(e)$) indicate the 
number of $C_4$'s and $K_4$'s that contain the vertex $i$ (resp.  edge $e$), respectively. 
The number of  diamonds, tailed triangles and $K_4$'s in the graph $G$ are 
denoted by $D(G)$, $TT(G)$, and $K_4(G),$ respectively. The number of wedges  
between two vertices $i$ and $j$ and ending at a vertex $i$ are written as 
$W(i,j)$ and $W(i).$ The numbers of the 5-motifs given in Figure~\ref{fig:escape5} are described with $N_i^5$, 
and of the ones in Figure~\ref{fig:6-motif} are described with $N_i.$ 

 A standard method for counting triangles is to enumerate the wedges and find the triangles by checking 
whether the missing edge is there or not. By a similar idea, the formulation here 
uses a cut-set, say $S$, for each motif $H$, 
whose removal disconnects $H.$ Let the components be $C_i$ and let $S\cup C_i$ be $H_i.$ 
There is some care needed in choosing 
this cut-set, however in our algorithm it is typically a vertex or an edge. 
The count for each possible $H_i$ that contains $S$ is obtained 
by the counts of 4-motifs and 5-motifs given in~\cite{escape}. The collection of 5-motifs that are used in our counting method can be seen in Figure~\ref{fig:escape5}. 
\subsection{Main Theorems}\label{main}
The exact computation for the motifs presented in Figure~\ref{fig:6-motif} is  
obtained in the following theorems.
We refer the reader to~\cite{escape} for 
 the technical details of the method used. 
 Here, we briefly discuss two examples to present  
 the general idea and how we apply it to obtain 
 Theorems~\ref{cut-v} and \ref{cut-e}. 
\begin{theorem}[Cut is a vertex]\label{cut-v}\\
$
\begin{array}{ll}
N_1 =&\sum_{i \in V}{{d(i)}\choose{5}}\\
N_2 =&\sum_{i \in V}{{W(i)}\choose{2}}(d(i)-2)
-N_7^5-N_2^5-2N_6^5-2TT(G) - 6D(G)\\
N_9 =&\sum_{i \in V}{T(i){{d(i)-2}\choose{3}}}\\
N_{11} =&\sum_{i \in V}{{T(i)}\choose{2}}(d(i)-4) - N_{11}^5\\
N_{12} =&\sum_{i\in V} K_4(i)(T(i)-3)-2N_{19}^5\\
N_{13} =&\sum_{i\in V} K_4(i){{d(i)-3}\choose{2}}\\
N_{14} =&\sum_{i\in V} K_4(i)(W(i)-6)-2N_{19}^5-2N_{15}^5\\
\end{array}
$
\end{theorem}
For example, the expression calculating $N_9$ in Theorem~\ref{cut-v} has no overcounting and it  considers every vertex as 
a cut-vertex $i$ and counts by 
pairing triangles and the three 
neighbors attached to $i$. 
\begin{figure}
     \centering
     \vspace{-.5cm}
     \includegraphics[scale=0.16]{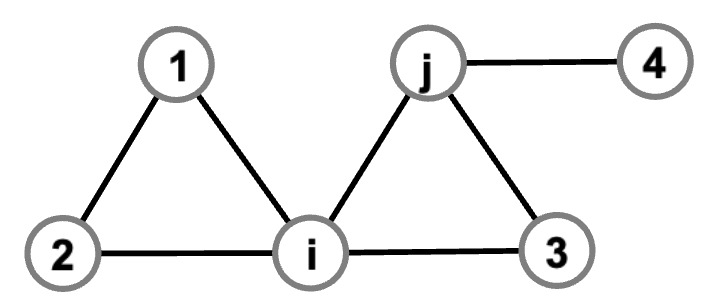}
     \caption{The chosen cut-edge for motif-17.}
     \vspace{-.5cm}
     \label{fig:motif17}
 \end{figure}
However, in the calculation of $N_{17}$, Theorem~\ref{cut-e}, 
we subtract the number of other motifs, counted unnecessarily. 
Here, the cut-set is an ordered pair $e=<i,j>$. 
One example of overcounting occurs 
when the vertices labeled 1 and 3 in Figure~\ref{fig:motif17} are 
chosen to be the same, meaning also 5-motifs with index 10 
are counted. Thus, we subtract it twice considering that  
$i$ is mapped to the vertices labeled either 1 or 4 (in Figure~\ref{fig:escape5}). Similarly, all subtractions 
remove the contributions of overcounting. 
\begin{theorem}[Cut is an edge]\label{cut-e}\\
Here, $<i,j>$ indicates an ordered and $(i,j)$ an unordered pair. \\
$N_3 =\sum_{(i,j) \in E} {{d(i)-1}\choose{2}}  {{d(j)-1}\choose{2}} - N_6^5 - D(G)$\\
$N_4 =\sum_{<i,j> \in E}{{d(i)-1}\choose{3}}(d(j)-1) - 2N_4^5$\\
$N_5 =\sum_{(i,j) \in E} {{d(j)-1}\choose{2}}(W(i)-2)-
-2N_7^{(5)}-2N_6^{(5)}-\sum_{x\in V}{{d(x)}\choose{4}}$\\
$N_6 =\sum_{(i, j) \in E}{[W(i)-(d(j)-1)][W(j)-(d(i)-1)]}-2N_4^5 - 5N_8^5 -2N_7^5 - 2N_5^5 - TT(G) - 3T(G)$\\
$N_7 =\sum_{e=(i, j) \in E}{(T(i)-T(e))(W(j)-(d(i)-1))}-2N_{12}^5-4N_9^5-8D(G)$\\
$N_8 =\sum_{e=<i,j> \in E}(T(i)-T(e)){{d(j)-1}\choose{2}}-2N_{11}^5 -6K_4(G)$\\
$N_{10} =\sum_{e=(i, j) \in E}(T(i)-T(e))(T(j)-T(e)) -N_{16}^5 - 6K_4(G)$\\
$N_{15} =\sum_{e\in E} {{K_4(e)}\choose{2}}-3N_{20}^5$\\
$N_{16} =\sum_{e\in E} K_4(e){{T(e)-2}\choose{2}}$\\
$N_{17} =\sum_{e=<i,j>\in E}(t(i)-t(e))t(e)(d(j)-2)-2N_{10}^5-12K_4(G)-2N_{16}^5$\\
\end{theorem}
\section{Experimental Results and Conclusions}\label{results}
 Experiments are performed on a computer that has 2.7 GHz dual-core Intel Core i5 processor, 3 MB L3 Cache and 8 GB 1867 MHz LPDDR3 memory. Our  counting formulas are implemented with C++ using ESCAPE \cite{escape} framework. The datasets are taken from \cite{network-repo,snap}. 

The input graph $G=G(V,E)$ is undirected and has $n$ vertices and 
$m$ edges, where multiple edges and loops are 
ignored. The input graph is stored as an adjacency list,
where each list is a hash table. Thus, edge queries
can be made in constant time. 
\begin{figure*}  
     \centering
     \vspace{-.5cm}
     \includegraphics[scale=0.3]{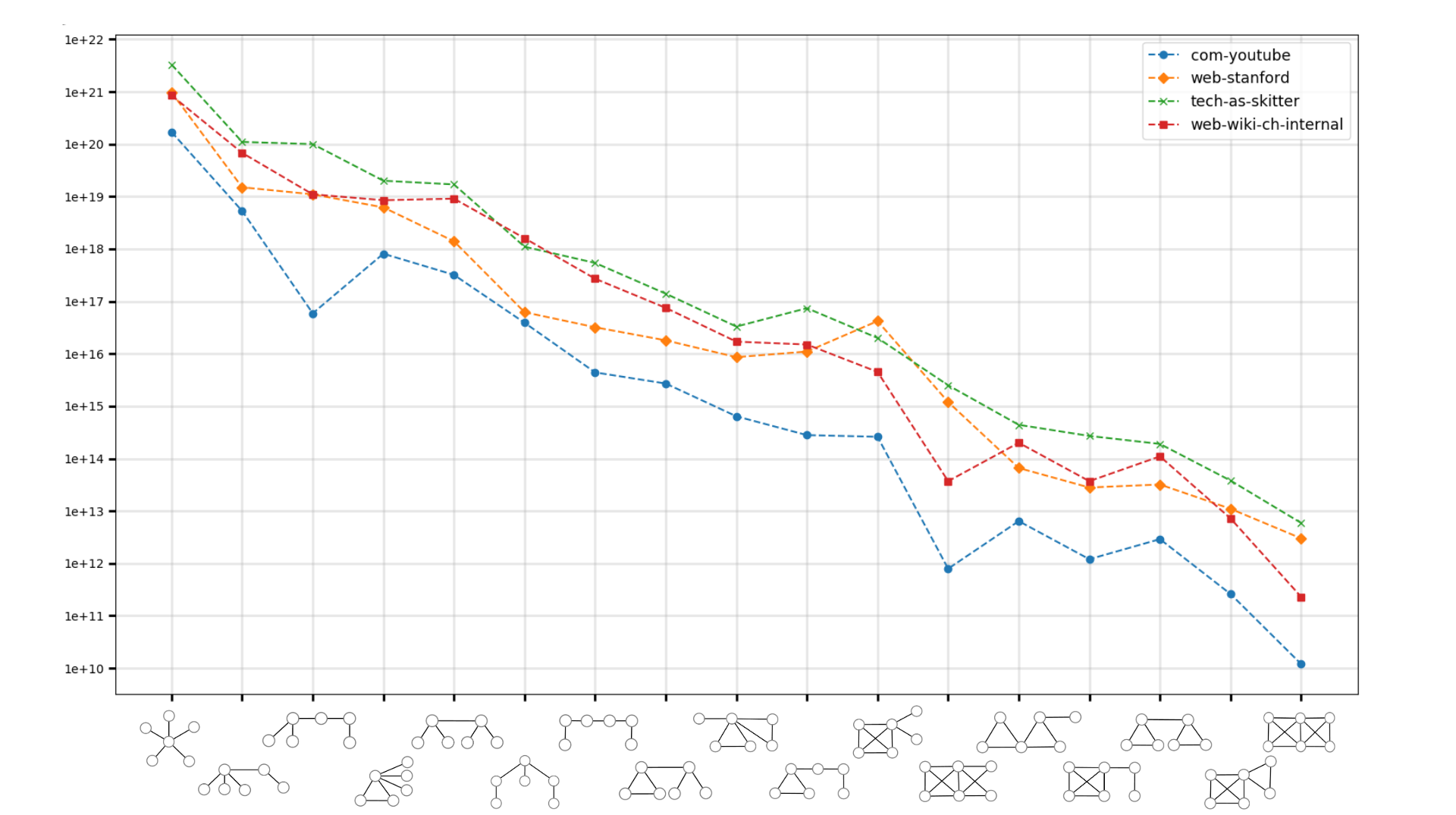}
     \caption{The counts of 6-motifs in the given networks.}
     \vspace{-.5cm}
     \label{fig:motif-distrib}
 \end{figure*}

The motifs studied here are not induced, however it is still possible to observe the behavior of the 
relationships in the corresponding network by the motif 
analysis obtained in Figure~\ref{fig:motif-distrib}. As expected, 
the most common motif is the 5-star and the tree motifs occur more frequently. One exception to that is the 5-star with an edge added.  This is not surprising, since this and the 5-star are two graphs,  abundant at the hub vertices with very high degrees in most social networks.  
Also, Figure~\ref{fig:motif-distrib} indicates that when the clique number of a motif is higher, the count of that motif is less. 

\begin{figure}
     \centering
    \vspace{-.4cm}
     \includegraphics[scale=0.3]{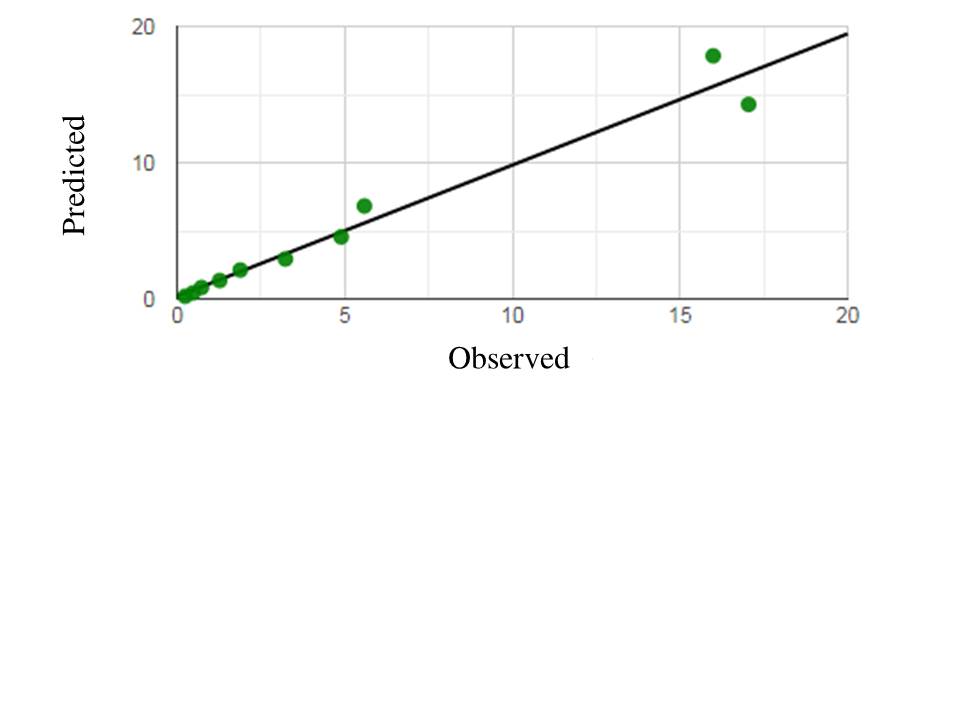}
      \vspace{-3cm}     
     \caption{The prediction of runtimes in seconds.} 
     \label{fig:predicted}
     \vspace{-.8cm}
 \end{figure}

In Table~\ref{table:dataset-motif-perf}, the 
runtimes of the algorithm together with the size of 
each network are provided. The fourth column shows 
the runtimes obtained in~\cite{escape} to 
evaluate the counts of motifs with 4 and 5 nodes. 
The runtime spent only for the 
counts of 6-motifs by our algorithm is listed in the last column. 
\begin{table}
\vspace{-0.5cm}
 \caption{The runtimes in seconds for the motif counts of various networks}
\begin{tabular}{|l|l|l|l|l|}
\hline
Network & $|V|$ & $|E|$& 4-5 motifs~\cite{escape} & 6-motif \\ \hline
com-youtube & 1.1M & 2.9M & 168.880 & 4.896 \\ \hline
web-wiki-ch-internal  & 1.9M & 8.9M & 2017.165 & 17.047 \\ \hline
web-stanford & 281.9K & 1.9M & 222.296 & 3.233 \\ \hline
tech-as-skitter & 1.7M   & 11.1M& 1401.271 & 15.991 \\ \hline
soc-brightkite & 56.7K & 212.9K & 6.629 & 0.242 \\ \hline
tech-RL-caida & 190.9K & 607.6K & 4.719 & 0.729 \\ \hline
flickr & 757.2K   & 1.4M & 13.008 & 1.886 \\ \hline
com-amazon & 334.8K   & 925.8K & 2.908 & 1.272 \\ \hline
web-google-dir & 875.5K   & 4.3M & 63.511 & 5.589 \\ \hline
ia-email-EU-dir & 265.0K   & 364.4K & 6.537 & 0.479 \\ \hline
\end{tabular}
\vspace{-.5cm}
\label{table:dataset-motif-perf}
\end{table}
The runtimes to count smaller motifs were 
predicted for any network in~\cite{escape}. 
Similarly, we obtain predictions using the runtimes 
in the last column of Table~\ref{table:dataset-motif-perf}, 
as shown in Figure~\ref{fig:predicted}. 
All counts in Theorems~\ref{cut-v} and \ref{cut-e} can be computed in time $O(n+m),$ where 
$n=|V|$ and $m=|E|.$ As social networks are sparse graphs and $|E|=O(n)$, 
our prediction is $-0.1476 + 1.6204 |E|$ seconds for any  network with $|E|$ edges. As observed in Table~\ref{table:dataset-motif-perf}, our algorithm is 
able to execute the counts of all 6-motifs in Figure~\ref{fig:6-motif} under 20 seconds, excluding the 
runtime spent to obtain the counts of smaller motifs. 
\subsection{Conclusions}
In this study, we presented a preliminary work that  
genaralizes the exact counting method for motifs of  networks in~\cite{escape} to a 
collection of 6-motifs with lower connectivity. 
We performed experiments to analyze the motif structure in real-world graphs and analyzed the runtime efficiency for the computations. 
The idea of counting 6-motifs by using algorithms based 
on the enumeration of smaller motifs results in much shorter runtime compared to other state-of-the-art algorithms. 
In a future study, we plan to extend this counting method to the remaining connected 6-motifs and use this data 
to obtain the counts of induced 6-motifs. 
\section*{Acknowledgements}
The research of the third author was supported in part by the BAGEP Award of the Science Academy of Turkey and by the TUBITAK Grant 11E283. 
\bibliographystyle{splncs04}
\bibliography{SHORT-T4-6motifs-LaleOzkahya}
\end{document}